\documentclass[12pt]{article}

\usepackage{setspace}
\usepackage{epsfig}
\def \ci {\cite}
\def\LB{\left(}
\def\RB{\right)}
\def\ba{\begin{eqnarray}}
\def\ea{\end{eqnarray}}
\def\de{\partial}
\def\ra{\rightarrow}
\def\pr{^{\prime}}

\def\mi{\mathrm{i}}

\def\Ll{\left<}

\def\Rr{\right>}
\def\Lb{\left|}
\def\Rb{\right|}
\def\vep{\varepsilon}
\def\nonum{\nonumber}
\def\bar{\begin{array}}
\def\ear{\end{array}}

\begin{document}
\enlargethispage*{1000pt}
\title{
Double Occupancy Errors in \\ Quantum Computing Operations: \\
Corrections to Adiabaticity}
\author{Ryan Requist$^{a,}$\footnote{rrequist@grad.physics.sunysb.edu} \
John Schliemann$^{b,}$\footnote{John.Schliemann@unibas.ch} \\
Alexander G. Abanov$^{a,}$\footnote{Alexandre.Abanov@stonybrook.edu} \
Daniel Loss$^{b,}$\footnote{Daniel.Loss@unibas.ch} \\
\small{\it $^a$Department of Physics and Astronomy} \\
\small{\it Stony Brook University} \\
\small{\it Stony Brook, NY 11794-3840}\\
\and
\small {\it $^b$Department of Physics and Astronomy} \\ 
\small {\it University of Basel} \\
\small {\it Klingelbergstrasse 82} \\
\small {\it CH-4056 Basel, Switzerland}}
\date{September 3, 2004}
\maketitle
\begin{abstract}
We study the corrections to adiabatic dynamics of two 
coupled quantum dot spin-qubits, each dot singly occupied 
with an electron, in the context of a quantum computing 
operation. Tunneling causes double occupancy at the 
conclusion of an operation and constitutes a processing
error.  We model the gate operation with an effective
two-level system, where non-adiabatic transitions 
correspond to double occupancy. 
The model is integrable and possesses three 
independent parameters. We confirm the accuracy of 
Dykhne's formula, a nonperturbative estimate of 
transitions, and discuss physically intuitive 
conditions for its validity. Our semiclassical results are in excellent
agreement with numerical simulations of the exact time evolution.
A similar approach applies to two-level systems in different contexts.

\vspace*{0.2cm}
\noindent
\scriptsize{
\begin{tabular}[b]{lcr} \parbox{6cm}{Keywords: Landau-Zener, 
non-adiabatic, tunneling, semiclassical, WKB, 
spin-qubits, quantum information processing} 
& &\parbox{6cm}{PACS numbers: 03.65.Sq, 
03.67.Lx, 73.23.Hk} \end{tabular}}
\end{abstract}
\thispagestyle{empty}

\hfill
\pagebreak

\newpage
\tableofcontents
\section{Introduction}
Quantum information processing is an active and fascinating 
direction of research with participation from various fields 
of physics and neighboring scientific disciplines~\cite{Nielsen00}. 
This extraordinary interest has generated a fairly vast amount 
of theoretical and experimental studies.  Possible experimental 
realizations of quantum information processing are presently 
being investigated. 
Among the different approaches, those in a solid state setting
are attractive, because they offer the potential of scalability 
--- the integration of a large number of quantum gates into 
a quantum computer once the individual gates and qubits are 
established.
With that in mind, several proposals for using electron and/or
nuclear spins in solid state systems have been put forward in 
recent years~\cite{Loss98,Privman98,Kane98,Barnes00,Levy01,Ladd01}.
Specifically, in Ref.~\cite{Loss98} it was proposed to use the 
spin of electrons residing in semiconductor quantum dots as qubits 
\cite{Burkard99,Hu00,Schliemann01a,Friesen02,Vandersypen02,Hanson03,Stepanenko03,review}.
In this paper we revisit the quantum dynamics of gate operations
between qubits of this type. Such two-qubit operations are performed 
by varying the amplitude of electron tunneling between the dots via 
external electric potentials. In a generic scenario, the tunneling 
amplitude between the dots is zero (or, more precisely, exponentially 
small) before and after the gate operation, while it is finite and 
appreciable during such a process. Thus, the typical time dependence 
of the tunneling amplitude is a pulse roughly characterized by its 
duration, amplitude, and ramp time (see Figure \ref{shape}.) 
During such a pulse, the tunneling amplitude is 
finite and essentially constant, and both electrons can explore the 
total two quantum dot system.
Therefore, their indistinguishable fermionic character is of relevance
\cite{Schliemann01a,Schliemann01b,Eckert02}. In 
particular, in such gate operations entanglement-like quantum 
correlations arise which require a description different from the 
usual entanglement between distinguishable parties (Alice, Bob,~...) 
in bipartite (or multipartite) systems. In such a case the proper 
statistics of the indistinguishable particles has to be taken into 
account \cite{Schliemann01a,Schliemann01b,Eckert02}.

Another important aspect of having a finite (as opposed to 
infinitely high) tunneling barrier between the dots is that it 
necessarily leads to (partially) doubly occupied states in the 
two-electron wave function, i.e. contributions to the wave function 
where both electrons are on the same dot (having different spin) occur 
with finite amplitude. 
Doubly occupied states which arise 
as the result of a measurement {\em after} 
the gate operation, destroy the information in those
qubits and lead to errors in the information processing. 
Therefore, it is desirable to reduce the probability of such errors, 
i.e. the occurrence of doubly occupied states, in the resulting 
two-electron state  {\em after} the gate operation, while it is 
necessarily finite during the operation \cite{Hu00,Schliemann01a}. 
If the error probability can be sufficiently reduced, error events
can be tolerable and handled with quantum error correction schemes.
An effective way of guaranteeing error suppression is to maintain
nearly adiabatic time evolution. Doubly occupied states 
then correspond to corrections to adiabatic evolution, which are 
often called ``non-adiabatic transitions''. 
Numerical simulations \cite{Schliemann01a} have shown that the 
adiabatic region, in terms of the pulse parameters such as ramp 
time and amplitude, is rather large. On a heuristic level, this 
numerical result is plausible on the basis of the classic papers on 
adiabatic quantum motion in two-level systems by Landau~\cite{Landau32}, 
Zener~\cite{Zener32}, Stueckelberg~\cite{Stuck32}, and Rosen and 
Zener~\cite{Rosen32}. For an overview see Ref.~\cite{Stenholm}.

In this work we study the quantum dynamics of the two-qubit gate 
operations described above and use Dykhne's semiclassical result to 
estimate the probability of non-adiabatic transition~\cite{Dykhne62}. 
The applicability of Dykhne's formula is analyzed from the standpoint 
of the theory of semiclassical approximations.  These semiclassical
estimates are found to be in excellent
agreement with numerical simulations of the exact time evolution.
Moreover, in a certain limit our model is integrable,
allowing us to explicitly calculate and interpret the corrections to 
Dykhne's formula. 

This paper is organized as follows. Section 2 reviews the 
derivation~\ci{Schliemann01a} of an effective two-level 
model. In section 3, we present our main result --- the asymptotic 
estimate of double occupancy, which in section 4 is compared with an 
integrable model and a numerical integration of the Schr\"{o}dinger 
equation.  In the Appendix, we construct the scattering matrix for the
integrable model, which has three independent parameters.

\section{Mapping to an Effective Two-level System}

For the purpose of studying double occupancy it is practical to
examine the dynamics of the quantum gate operation in a subspace 
spanned by singly and doubly occupied states. Following 
Ref.~\ci{Schliemann01a} with only minor changes of notation, we now detail 
how to reduce the description of a system of 
two coupled quantum dot spin qubits to an effective two-level Hamiltonian.
The system is described by a Hamiltonian of the form 
${\cal H}=T+C$, where $C$ denotes the Coulomb
repulsion between the electrons, and $T=\sum_{i=1,2}h_{i}$ is the
single-particle part with
\begin{equation}
h_{i}=\frac{1}{2m}\left(\vec p_{i}+\frac{e}{c}\vec A(\vec r_{i})\right)^{2}
+V(\vec r_{i})\,.
\label{opham}
\end{equation}
The single-particle Hamiltonian $h_{i}$ describes electron dynamics confined
to the $xy$-plane in a perpendicular magnetic field $\vec{B}$. The 
effective mass $m$ is a material dependent parameter.
The coupling of the dots (which includes tunneling)
is modeled by a quartic potential
\begin{equation}
V(\vec r)=V(x,y)=\frac{m\omega^{2}_{0}}{2}
\left(\frac{1}{4a^2}\left(x^{2}-a^{2}\right)^{2} +y^{2}\right)\,,
\end{equation}
which separates into two harmonic wells of frequency $\omega_{0}$ (one for
each dot) in the limit $a\gg a_{0}$, where $a$ is half the distance between
the dots and $a_{0}=\sqrt{\hbar/m\omega_{0}}$ is the effective Bohr radius
of a dot.

Following Burkard {\it et al.} \cite{Burkard99} we employ the Hund-Mulliken
method of molecular orbits to describe the low-lying spectrum of our system.
This approach concentrates on the lowest orbital states in each dot and is
an extension of the Heitler-London method \cite{Burkard99}.
The Hund-Mulliken approach accounts for
the fact that both electrons can, in the presence of a finite tunneling
amplitude, explore the entire system of the two dots, and therefore adequately
includes the possibility of doubly occupied states. 
In the usual symmetric gauge $\vec A=B(-y,x,0)/2$ the Fock-Darwin ground state
of a single dot with harmonic confinement centered around $\vec r=(\pm a,0,0)$
reads
\begin{eqnarray}
\varphi_{\pm a}(x,y) & = & \sqrt{\frac{m\omega}{\pi\hbar}}
\exp\left(-\frac{m\omega}{2\hbar}\left(\left(x\mp a\right)^{2}+y^{2}\right)
\right)\nonumber\\
 & & \cdot\exp\left(\mp\frac{i}{2}y\frac{a}{l_{B}^{2}}\right)\,,
\label{dotstates}
\end{eqnarray}
where $l_{B}=\sqrt{\hbar c/eB}$ is the magnetic length, and the frequency
$\omega$ is given by $\omega^{2}=\omega_{0}^{2}+\LB \omega_{L}/2 \RB^{2}$ where
$\omega_{L}=eB/mc$ is the usual Larmor frequency. From these non-orthogonal
single-particle states we construct the orthonormalized states $|A\rangle$ and
$|B\rangle$ with wave functions
\begin{eqnarray}
\langle\vec r|A\rangle & = &
\frac{1}{\sqrt{1-2Sg+g^{2}}}\left(\varphi_{+a}-g\varphi_{-a}\right)\,,\\
\langle\vec r|B\rangle & = &
\frac{1}{\sqrt{1-2Sg+g^{2}}}\left(\varphi_{-a}-g\varphi_{+a}\right)\,,
\end{eqnarray}
with $S$ being the overlap between the states (\ref{dotstates}) and
$g=(1-\sqrt{1-S^{2}})/S$.
For appropriate values of system parameters such as the interdot distance
and the external magnetic field, the overlap $S$ becomes exponentially small
\cite{Burkard99}. In this limit an electron in one of the states
$|A\rangle$, $|B\rangle$ is predominantly localized around
$\vec r=(\pm a,0,0)$.
In the following we consider this case and use these states as basis states
to define qubits, i.e.
qubits are realized by the spin state of an electron in either orbital
$|A\rangle$, or orbital $|B\rangle$.

An appropriate basis set for the six-dimensional two-particle Hilbert space is
given (using standard notation) by the three spin singlets
\begin{eqnarray}
|S_{1}\rangle & = & \frac{1}{\sqrt{2}}
\left(c^{+}_{A\uparrow}c^{+}_{B\downarrow}-
c^{+}_{A\downarrow}c^{+}_{B\uparrow}\right)|0\rangle\,,\\
|S_{2}\rangle & = & \frac{1}{\sqrt{2}}
\left(c^{+}_{A\uparrow}c^{+}_{A\downarrow}+
c^{+}_{B\uparrow}c^{+}_{B\downarrow}\right)|0\rangle\,,\\
|S_{3}\rangle & = & \frac{1}{\sqrt{2}}
\left(c^{+}_{A\uparrow}c^{+}_{A\downarrow}-
c^{+}_{B\uparrow}c^{+}_{B\downarrow}\right)|0\rangle\,,
\end{eqnarray}
and the triplet multiplet,
\begin{eqnarray}
|T^{-1}\rangle & = &
c^{+}_{A\downarrow}c^{+}_{B\downarrow}|0\rangle\,,\\
|T^{0}\rangle & = & \frac{1}{\sqrt{2}}
\left(c^{+}_{A\uparrow}c^{+}_{B\downarrow}+
c^{+}_{A\downarrow}c^{+}_{B\uparrow}\right)|0\rangle\,,\\
|T^{1}\rangle & = & c^{+}_{A\uparrow}c^{+}_{B\uparrow}|0\rangle\,.
\end{eqnarray}
As the Hamiltonian conserves spin, the three triplet states are 
degenerate eigenstates (typically we can ignore possible
Zeeman splittings \cite{Burkard99}) and have the
eigenvalue,
\begin{equation}
\varepsilon_{\mathrm{trip}}=2\varepsilon_1+V_{-}\,,
\end{equation}
where we have defined
\begin{equation}
\varepsilon_1 = \langle A|h_1|A\rangle=\langle B|h_1|B\rangle
\end{equation}
and the expectation value of Coulomb energy
\begin{equation}
V_{-}=\langle T^{\alpha}|C|T^{\alpha}\rangle\quad,\quad
V_{+}=\langle S_{1}|C|S_{1}\rangle\,.
\end{equation}

An important further observation is that, as a consequence of inversion
symmetry along the axis connecting the dots,
the Hamiltonian does not
have any non-zero matrix elements between the singlet state $|S_{3}\rangle$
and other states. Hence, $|S_{3}\rangle$ is, independently of
the system parameters, an eigenstate. The eigenvalues of the triplet and
$|S_{3}\rangle$, however, do depend on system parameters. The
Hamiltonian acting on the remaining space spanned by $|S_{1}\rangle$ and
$|S_{2}\rangle$ can be written as
\begin{equation}
{\cal H}=2\varepsilon_1+\frac{1}{2}U_{H}+V_{+}-
\frac{U_H}{2} \left(
\begin{array}{cc}
 1 & t_{H} \\ t_{H} & -1
\end{array}\right)
\label{2ham}
\end{equation}
where
\begin{equation}
t_{H} = -\frac{4}{U_H} \LB \langle A|h_1|B\rangle+\frac{1}{2}\langle S_{2}|C|S_{1}\rangle \RB
\end{equation}
and
\begin{equation}
U_{H}=\langle S_{2}|C|S_{2}\rangle-V_{+}\,.
\end{equation}
The nontrivial part of (\ref{2ham}) is a simple Hubbard Hamiltonian
on two sites and can be identified as the Hamiltonian of a
pseudospin-$\frac{1}{2}$ object in a
pseudomagnetic field having a component $U_{H}$ in the $\hat z$-direction 
and $U_H t_{H}$ in the $\hat x$-direction of pseudospin space.  [Note 
that this pseudospin is not related to the spin degree of freedom which 
constitutes the qubit!]  The basis states
themselves are eigenstates only in the case of vanishing tunneling 
amplitude $t_{H}$ where $|S_{1}\rangle$ is the ground state and 
$|S_{2}\rangle$ is a higher lying state due of the Coulomb (Hubbard) 
energy. In all other cases, the ground state has an admixture of 
doubly occupied states contained in $|S_{2}\rangle$.
The energy gap between the triplet and the singlet ground state is
\begin{equation}
\varepsilon_{\mathrm{trip}}-\varepsilon_{\mathrm{gs}}=V_{-}-V_{+}
-\frac{U_{H}}{2}+\frac{U_H}{2}\sqrt{1+t^{2}_{H}}\,.
\label{stsplitting}
\end{equation}

A key challenge for state-of-the-art quantum information processing 
is the construction of systems composed of two coupled quantum dots 
which can be coupled to perform
swap operations ${\cal U}_{SW}$, i.e. unitary two-qubit operations
which interchange the spin states (qubits) of the
electrons on the two dots.
By combining the ``square root'' ${\cal U}_{SW}^{1/2}$
of such a swap with other isolated-qubit manipulations one can construct a 
quantum
XOR gate. A quantum XOR gate, along with isolated-qubit operations, has been
shown to be sufficient for the implementation of any quantum algorithm
\cite{DiVincenzo95}. Hence
a practical and reliable realization of a swap gate would be an important
step toward the fabrication of a solid state quantum computer.
A swap operation in the present system is a unitary transformation which turns
a state having the qubits in different states, say,
\begin{equation}
c^{+}_{A\uparrow}c^{+}_{B\downarrow}|0\rangle=
\frac{1}{\sqrt{2}}\left(|T^{0}\rangle+|S_{1}\rangle\right)\,,
\label{instate}
\end{equation}
into a state where the contents of the qubits are interchanged,
\begin{equation}
c^{+}_{A\downarrow}c^{+}_{B\uparrow}|0\rangle=
\frac{1}{\sqrt{2}}\left(|T^{0}\rangle-|S_{1}\rangle\right)\,.
\label{outstate}
\end{equation}
These two states are eigenstates in the case $V_{+}=V_{-}$ and $t_{H}=0$
for which the singlet-triplet splitting vanishes.

As discussed in references \cite{Loss98,Burkard99,Schliemann01a},
swapping may be achieved
by the action of a gate that lowers the potential barrier between the
quantum dots.  This
leads to exponentially larger values for both $V_{+}-V_{-}$ and
$t_H$.  It is adequate for our purposes to consider a model where
$V_{+}=V_{-}$ (consistent with the above limit of small overlap $S$), and
the singlet-triplet splitting results entirely from
$t_H$.  If the duration and amplitude of a tunneling pulse
is adjusted
appropriately, the relative dynamical phase between the singlet and 
the triplet state accumulates a shift of $\pi$, 
\ba
\int_{-\infty}^{\infty} \; \mathrm{d}t \LB \vep_{\mathrm{trip}}(t) 
- \vep_{\mathrm{gs}}(t) \RB
= \pi \label{constraint}
\ea
and the swapping operation between states (\ref{instate}) and 
(\ref{outstate}) is performed. However, during the operation
the state $\Lb S_1 \Rr$ is coupled to $\Lb S_2 \Rr$, and they evolve
according to (\ref{2ham}).  Double occupancy errors are thus generically
introduced.

The reduction of the dynamics to the time evolution of a two-level system
relies on the fact that the system has inversion symmetry along the
$\hat x$-axis in real space connecting the dots. This symmetry can be
broken if odd powers of the particle coordinates $x_{i}$ are added to
the Hamiltonian (\ref{opham}), for example the potential of a
homogeneous electric field. The breaking of inversion symmetry introduces
additional matrix elements between $|S_{3}\rangle$ and the other two 
singlets leading to an effective three level Hamiltonian. However, as it 
was shown in Ref.~\cite{Schliemann01a}, this more inclusive Hamiltonian has 
qualitatively the same properties concerning non-adiabatic dynamics 
as the two-level system on which we shall concentrate in the following.

So far we have not considered a possible Zeeman coupling to the electron spin.
This would not change the situation essentially since all states
involved in the
swapping  process ($|T^{0}\rangle$, $|S_{1}\rangle$, $|S_{2}\rangle$, and
possibly $|S_{3}\rangle$) have the
total spin quantum number $S^{z}=0$.

\section{Analysis of Non-adiabatic Transitions}

In this section we use Dykhne's formula for non-adiabatic 
transitions to derive an asymptotic expression for 
the probability of final double occupancy, given physically 
motivated properties of the two-qubit operation. 

\begin{figure}[htb]
\begin{center}
\includegraphics[width=0.5\textwidth]{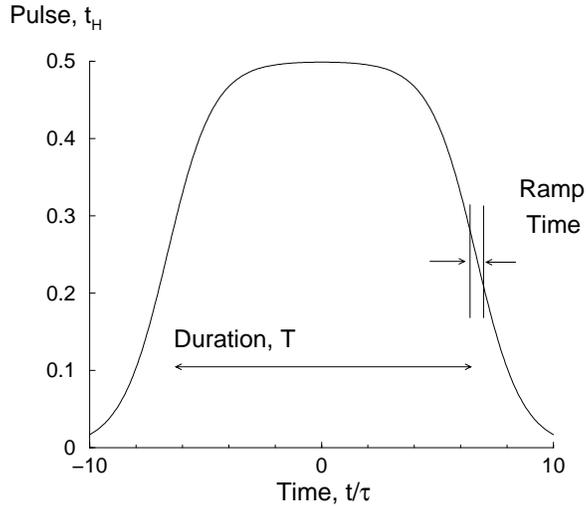}
\caption{A realistic profile of the tunneling pulse, 
labeled with the characteristic time scales.
Corresponding to  $\lambda=2$ and $\delta=\frac{1}{2}$,
the time scales shown are $\tau$ and 
$T \approx 13 \tau$.}
\label{shape}
\end{center}
\end{figure}

\begin{figure}[htb!]
\begin{center}
\includegraphics[width=0.5\textwidth]{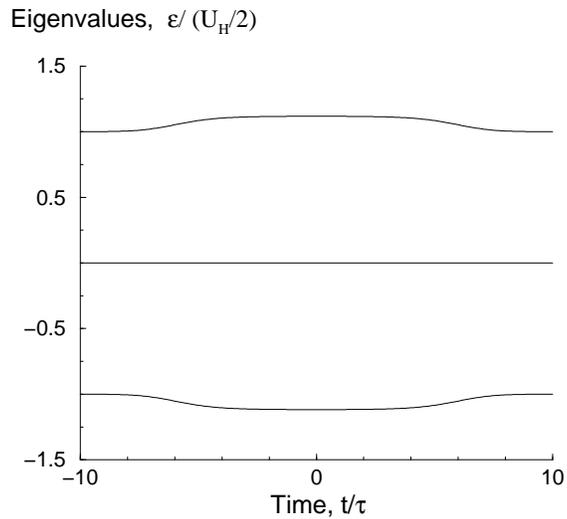}
\caption{A profile of the eigenvalues $\pm \vep(t)$
corresponding to the pulse of Figure \ref{shape}.}
\label{energy}
\end{center}
\end{figure}

As described in the previous section, the modulation of the
tunneling barrier during the swapping process induces a 
coupling between the singly occupied qubit state 
$\left| S_1 \right>$ and the doubly occupied state 
$\left| S_2 \right>$.  Their dynamics are 
governed by the effective Hamiltonian 

\ba
\mathcal{H}_{\mathrm{eff}} = - \frac{U_H}{2} 
\left( \begin{array}{cc} 1 & t_H \\
 t_H & -1 \end{array}\right)  \label{H1}
\ea
in the $\left| S_{1,2} \Rr$ basis.
The terms omitted from (\ref{2ham}) do not contribute to transitions, 
because the identity operator in the $\left| S_{1,2} \Rr$ basis 
commutes with the remainder of the Hamiltonian.
The large energy offset $U_H$ between singly and doubly occupied states, 
primarily due to the Coulomb repulsion, is perturbed only by
an exponentially small additive quantity (proportional to the overlap, $S$) 
during the swapping operation and is hereafter assumed to be 
a constant. Our specification of the pulse (Figure \ref{shape})
\ba
t_H(t) = \frac{\delta}{1+\frac{\cosh(t/\tau)}{\cosh(T/2\tau)}} \label{pulse}
\ea
with dimensionless strength $\delta$ is considered to 
realistically reflect the tunneling
amplitude that would arise from a modulation of the gate
potential~\ci{Schliemann01a}. The exponential dependence of the
ramping near $t=\pm T/2$ has its origin in the exponential 
sensitivity of the coupling to the gate voltage and in turn the 
exponential decay 
of the single-particle wavefunctions (\ref{dotstates}) in the 
inter-dot region~\ci{Burk}.
The pulse mimics a step of duration $T$ and magnitude $\delta U_H/2$,
whose ramping on and off have a characteristic time 
$\tau$. The perturbation of the instantaneous 
eigenvalues by the pulse is shown in Figure 
\ref{energy}.


The Schr\"{o}dinger equation is 
\ba
\mi\,\hbar \, \frac{\mathrm{d}}{\mathrm{d}t} \; \Lb\psi(t) \Rr = 
\mathcal{H}_{\mathrm{eff}}(t) \; \Lb\psi(t) \Rr .
\ea
Our task is to 
find the component of double occupancy in the final state,
$\Ll S_2 | \psi(\infty) \Rr$, given that the prepared state is purely
singly occupied, $\left| \Ll S_1 | \psi(-\infty) \Rr \right| =1$.  

Our model involves three dimensionless scales, assigned for 
our purposes as follows: $\delta$, $\lambda\equiv U_H \tau/2\hbar$, 
and $\eta = T/\tau$. 
Presently, the case of interest is 
\ba
\lambda \gg 1 \qquad \mathrm{and} \qquad \eta \gg 1.  \label{lims}
\ea 
The first of these conditions reflects the adiabaticity of the 
problem.  The second requires that the ramping on and ramping
off of the pulse be temporally well-separated and distinct events.

Let us pause and for this paragraph review the familiar notions of 
transitions 
under the action of a time dependent perturbation. The pulse acts as 
a transient perturbation and otherwise the Hamiltonian (\ref{H1}) is 
diagonal. By force of the adiabatic theorem, the probability of 
transition among eigenstates vanishes in the limit $\tau \ra \infty$,
where the ramping on and off of the pulse is adiabatic. 
In the zeroth order of adiabatic perturbation theory, there are no 
transitions, and the leading behavior of the general solution is simply 
the dynamical phase of each component eigenstate
\ba
\Lb\psi(t) \Rr &\approx& 
e^{\,\frac{\mi}{\hbar} \int^t_{-\infty} \; \mathrm{d}t\pr \vep(t\pr)} 
\; \Lb \xi_1(t) \Rr \Ll \xi_1(-\infty)  | \psi(-\infty) \right>  \nonum \\
& +& e^{-\,\;\frac{\mi}{\hbar} \int^t_{-\infty} \; \mathrm{d}t\pr \vep(t\pr)} 
\;\; \Lb \xi_2(t) \Rr \Ll \xi_2(-\infty)  | \psi(-\infty) \right> ,
\label{zeroth}
\ea
where $\Lb \xi_{1,2}(t) \Rr$ are the instantaneous eigenstates 
(given explicitly in (\ref{estate})) of 
Hamiltonian (\ref{H1}) corresponding to eigenvalues 
\ba
\mp\vep(t) = \mp\frac{U_H}{2} \sqrt{1 + t^2_H}\;\;, 
\label{eval}
\ea
respectively. 
In general, the zeroth order approximation should also include a 
factor representing Berry's phase.  However, for a 
real symmetric Hamiltonian such as (\ref{H1}), Berry's phase is
irrelevant, because the Hamiltonian has an inherent planarity.
In pseudospin one-half notation, 
$\mathcal{H}_{\mathrm{eff}} = \vec H \cdot \vec \sigma$,
the time evolution of the pseudomagnetic field $\vec{H} = \vec{H}(t)$ 
is in a plane. If the azimuthal axis (north pole) is chosen to lie 
within that plane, the solid angle subtended by the pseudomagnetic 
field vanishes identically. Although Berry's phase is out 
of consideration, there are interesting circumstances where 
Berry's phase is relevant to transitions. It can correct
the transition amplitude \cite{Berry91} and produce 
topological selection rules for spin tunneling \cite{Loss92,Loss}. 
Our problem is one of a class initiated by 
the work of Landau, Zener, and Stueckelberg (LSZ) 
\cite{Landau32,Zener32,Stuck32}. 
However, we emphasize that for our model (with the pulse specified as
(\ref{pulse})) the linearization
of Hamiltonian matrix elements near the times where adiabaticity
is most severely violated is not
applicable and leads to an incorrect result. As we will see the
shape of the pulse is important.

\subsection{Application of Dykhne's Formula}

\begin{figure}[htb]
\begin{center}
\includegraphics[width=0.8\textwidth]{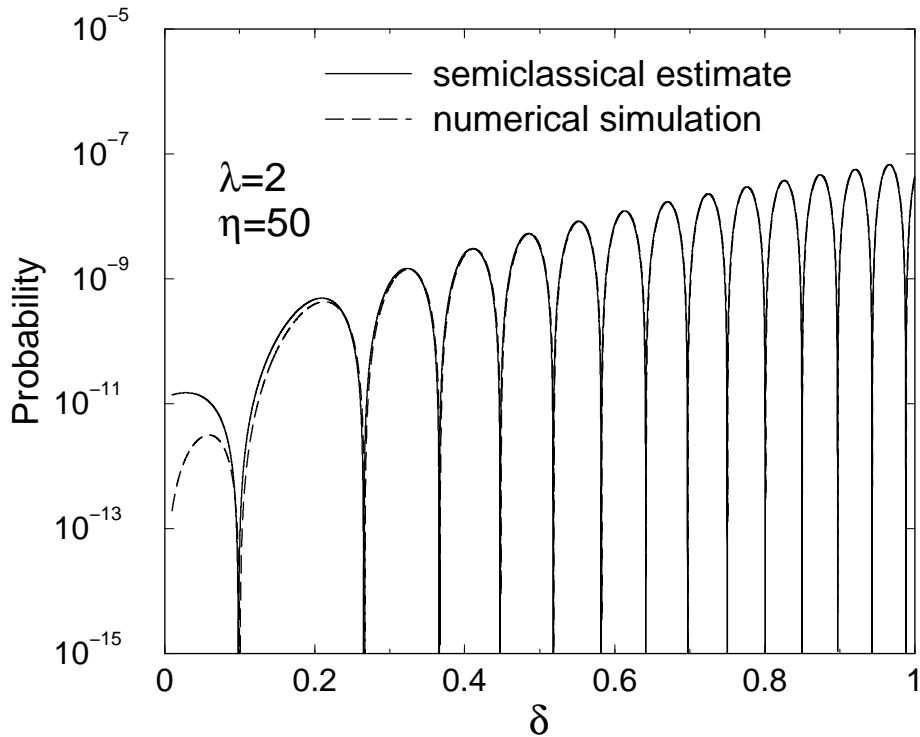}
\caption{The probability for non-adiabatic transitions for $\lambda=2$ and
$\eta=50$ as a function of $\delta$. We compare our semiclassical estimate
according to expression (\ref{main}) with results from numerical
simulations of the exact quantum mechanical time evolution 
as done in Ref.~\cite{Schliemann01a}. The results are in excellent
agreement.
\label{numerical1}}
\end{center}
\end{figure}

\begin{figure}[htb]
\begin{center}
\includegraphics[width=0.8\textwidth]{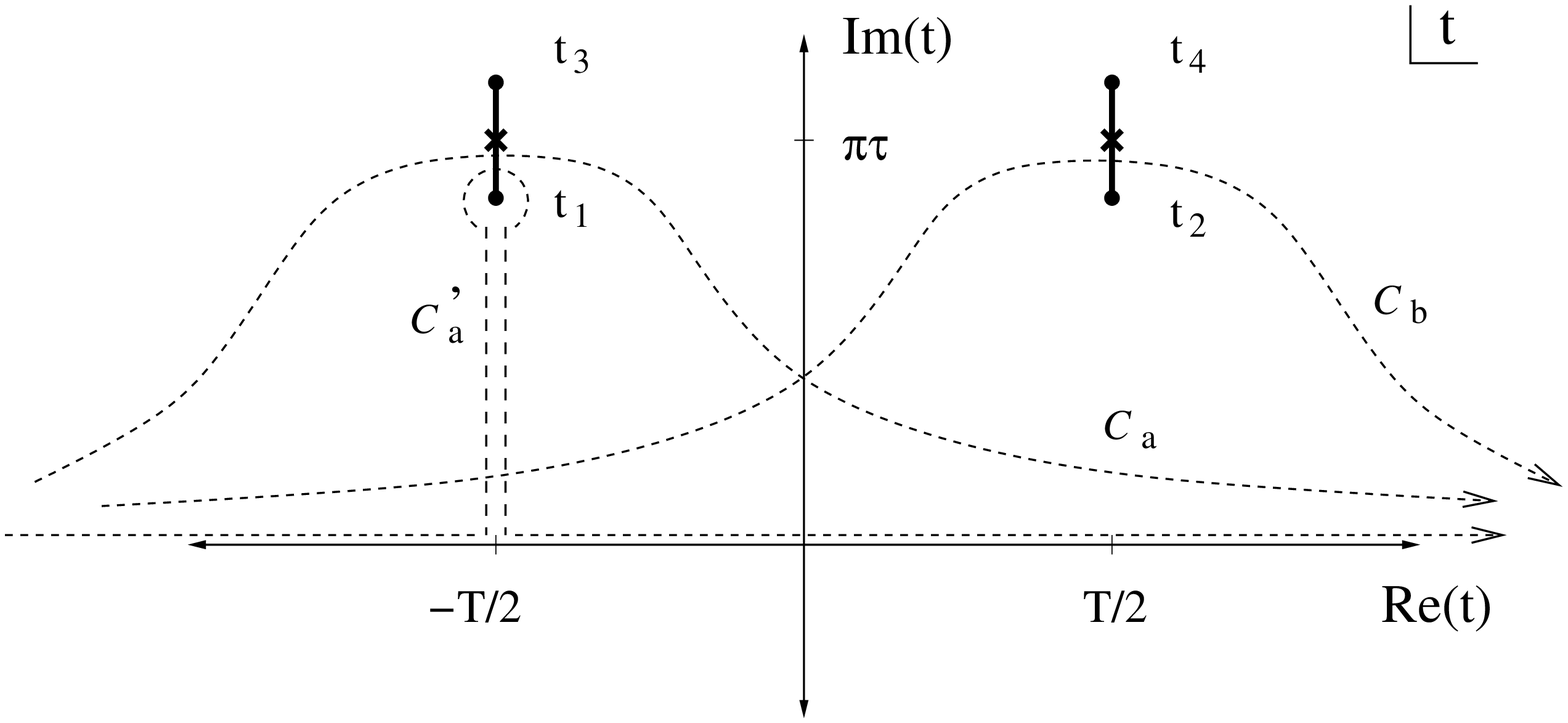}
\caption{The analytic structure of the function
 $\vep=\vep(t)$ shown only in a segment of the upper
half plane. The contour $\mathcal{C}_a$ is associated
with transitions that occur due to the ramping on of 
the pulse, while contour $\mathcal{C}_b$ is associated
with the ramping off.
By Cauchy's theorem, an integral on the contour 
$\mathcal{C}_a$ is equal to the integral on the 
contour $\mathcal{C}_a^{\prime}$.
Bold lines represent 
branch cuts, dots represent branch points, and 
poles are denoted with a $\times$.}
\label{contour}
\end{center}
\end{figure}

Returning to our model, we observe that if the time interval 
$t\in(-\infty, \infty)$ is divided into two domains $t<0$ and $t>0$, 
and in the limit $\eta \equiv \frac{T}{\tau} \gg 1$, 
the pulse (\ref{pulse}) is approximated by
\ba
t_H(t) \approx \left\{ \begin{array}{ll} 
  \frac{\delta}{1+ e^{-\frac{t}{\tau}-\frac{T}{2\tau}}} & \qquad t<0 \\[0.5cm]
  \frac{\delta}{1+ e^{\frac{t}{\tau}-\frac{T}{2\tau}}} & \qquad t>0.
\end{array} \right. \label{time}
\ea
In each domain the pulse behaves as a step,  
and the dynamics are integrable (see Section 4.) 
We will focus first on the interval $t<0$, where the 
probability of transition to a doubly occupied state $P_<$ may be 
estimated with Dykhne's formula~\cite{Dykhne62} 
\ba
P_{<} =  \Lb \Ll S_2 | \psi(0) \Rr \Rb^2 
\sim \; 
e^{-\frac{4}{\hbar} \;\Im \int_{\Re(t_1)}^{t_1} \; \mathrm{d}z \;\vep(z)},
\label{Dykhne}
\ea
where the approximation (\ref{time}) is used implicitly  
for the instantaneous eigenenergies $\mp \vep(t)$ 
defined above by (\ref{eval}).
The turning point $t=t_1$, given explicitly below, is a 
{\em complex root} of the function $\vep=\vep(t)$; in other words, it is 
an intersection of the energy surfaces (curves) of the 
two instantaneous (``frozen'') eigenstates.
Our model is the patching together of two domains of time, 
and transitions that occur during $t<0$ and
$t>0$ interfere.  The expression for the probability of 
transition during the time evolution from $t=-\infty$ to $t=\infty$ is
\ba
P &=&  \left| \Ll S_2 | \psi(\infty) \Rr \right|^2 \nonum
\sim \Lb e^{\;\frac{\mi}{\hbar}  \int_{\mathcal{C}_a} \mathrm{d}z \; \vep(z)} 
+ e^{\;\frac{\mi}{\hbar} \int_{\mathcal{C}_b} \mathrm{d}z \; \vep(z)} \Rb^2 \\[0.5cm]
&\sim& \;  
\left| 
e^{\;\frac{\mi}{\hbar} \Re \int_{\mathcal{C}_a} \mathrm{d}z \; \vep(z)}
e^{\;-\frac{2}{\hbar} \;\Im \int_{\Re(t_1)}^{t_1} \; \mathrm{d}z \;\vep(z)} 
+ e^{\;\frac{\mi}{\hbar} \Re \int_{\mathcal{C}_b} \mathrm{d}z \; \vep(z)}
e^{\;-\frac{2}{\hbar} \; \Im \int_{\Re(t_2)}^{t_2} \; \mathrm{d}z \;\vep(z)}
\right|^2 \label{Dyk1}\\[0.5cm]
&=& 4\; \sin^2 \LB \frac{1}{\hbar} \;\Re \int_{t_1}^{t_2}\; \mathrm{d}z\; \vep(z) \RB 
 \;\; P_< , 
\label{Dyk2}
\ea
where the contours $\mathcal{C}_{a,b}$ are shown in Figure~\ref{contour},
and according to the sign of the integration variable, 
$\textrm{sgn} (\Re \;z)$,  
one or the other of the approximations (\ref{time}) is used.  
The turning points $t=t_{1,2}$ appearing in the limits of 
integration of (\ref{Dyk1}), are chosen as the two roots
of $\vep=\vep(t)$ that are closest to and above the real 
time axis (see Figure~\ref{contour})
\ba
t_{1,2} = \mp \LB \frac{T}{2} + \tau \; \ln(\sqrt{1+\delta^2}) \RB 
+ \mi \;\tau \LB \pi - \arctan(\delta) \RB . \label{tp}
\ea 
They are nonreal because the Hamiltonian (\ref{H1}) is 
nondegenerate for real times. Equation (\ref{Dyk2}) follows from
(\ref{Dyk1}), because the symmetry of the pulse 
implies $\Im(t_1)=\Im(t_2)$ and $P_<=P_>$.
The oscillatory first factor of (\ref{Dyk2}) is the interference of 
the dynamical phase of each term of (\ref{Dyk1}). 
The magnitude of $P$ is dominated by
the second factor, $P_<$, whose 
exponent is given by the following integral 
\ba
&&\;\;-4 \lambda \;\Im\; \int_{\ln(\sqrt{1+\delta^2})}^{\ln(\sqrt{1+\delta^2})+\mi\pi - \mi \; \arctan(\delta)} 
\mathrm{d}z \LB 1+\LB \frac{\delta}{1+e^{z}}\RB^2 \RB^{-\frac{1}{2}} \nonum \\[0.5cm]
&=& \;-2\pi \lambda \LB 1 + \sqrt{1+\delta^2} - \delta \RB .
\ea
Substituting this result in (\ref{Dykhne}) we have
\ba
P_< \sim 
e^{-2\pi \lambda \left( 1+ \sqrt{1+\delta^2}- \delta \right)}.
\label{main2}
\ea
From (\ref{Dyk2}), we have our main result, 
an asymptotic estimate for the probability of final double-occupancy 
\ba
P  \sim
4 \sin^2 
\LB \frac{1}{\hbar} \;\Re \int_{t_1}^{t_2} \; \mathrm{d}z \; \vep(z) \RB 
\; e^{-2\pi \lambda \left( 1+ \sqrt{1+\delta^2}- \delta \right)} , \label{main}
\ea
which is shown as a function of $\delta$ in Figure~\ref{numerical1}.
The probability $P$ is characteristically nonperturbative in 
the adiabatic limit $\tau \ra \infty$ with $U_H$ fixed, or 
equivalently $\lambda \ra \infty$. Hence, the dimensionless 
quantity associated with the exponential suppression is $\lambda$
and has been called the ``adiabaticity parameter.''
For $\eta \equiv \frac{T}{\tau} \gg 1$, the approximation (\ref{time}) 
allows us to estimate the argument of the prefactor of (\ref{main}) to
exponential accuracy, 
\ba
\frac{1}{\hbar} \;\Re \int_{t_1}^{t_2}\; \mathrm{d}z\; \vep(z)
&=& \sqrt{1+\delta^2} \;\lambda \;\eta 
- 2 \lambda\, \left\{ \; \ln \LB \sqrt{1+\delta^2} + 1 \RB \right.
\nonum\\
& & \mbox{} - \sqrt{1+\delta^2}\; \ln \LB 2(1+\delta^2) \RB 
+ \delta\; \ln \LB \sqrt{1+\delta^2} + \delta \RB 
\nonum\\
& & \mbox{} + \left. \LB \sqrt{1+\delta^2} -1 \RB \; \ln(\delta)\; \right\}
+\; \mathcal{O}\LB e^{-\eta/2}\RB . \label{argument}
\ea
The oscillation with respect to the duration of the pulse $T$ 
is reminiscent of a similar 
factor in the Rosen-Zener model.  The phenomenon of pulsed 
perturbations that return the full amplitude/occupation to the 
initial state have been studied in the context of atom-laser 
interactions~\cite{Bambini81,Robinson84,Vitanov94,Robinson95}.
\begin{figure}[htb]
\begin{center}
\includegraphics[width=0.8\textwidth]{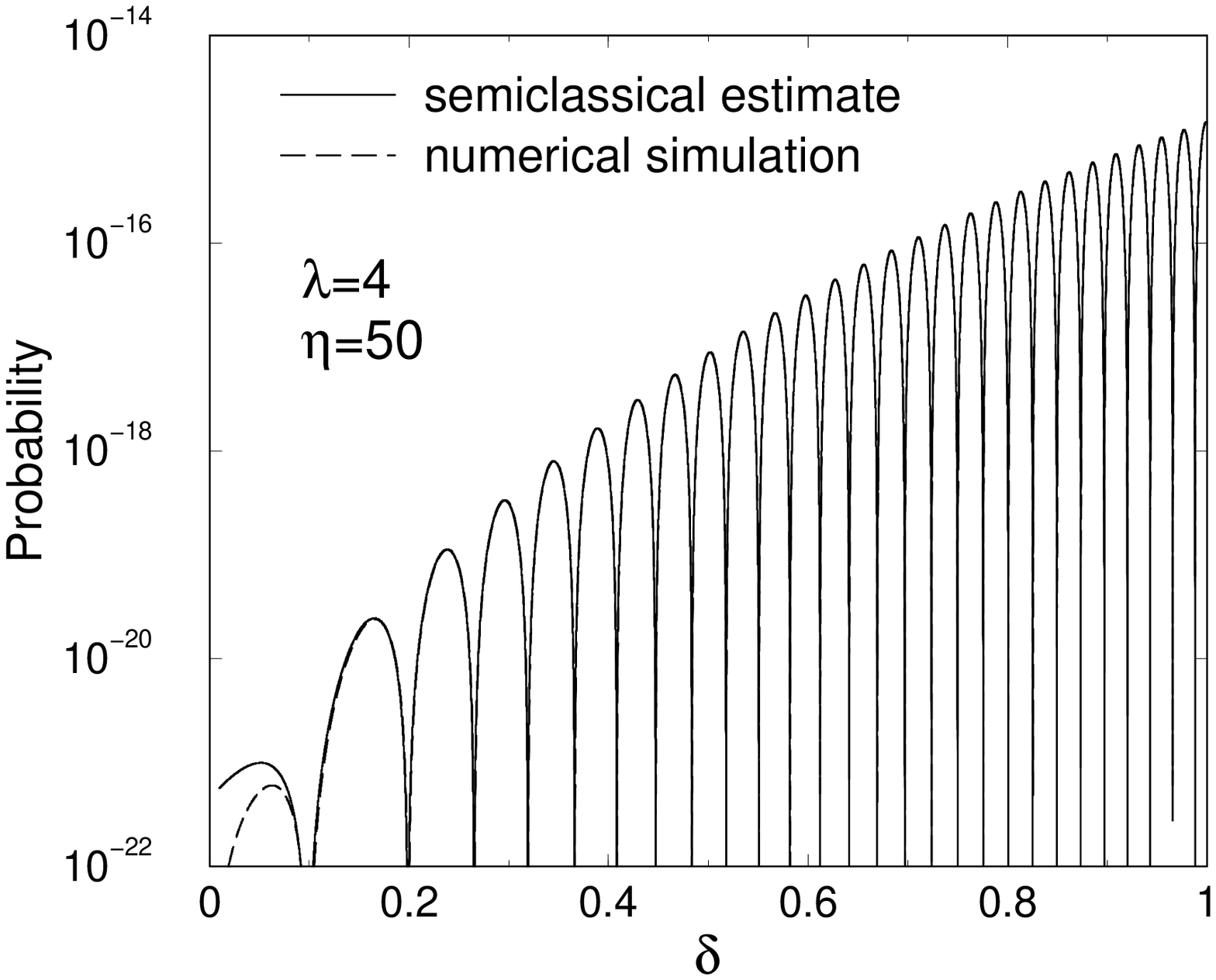}
\caption{The probability for non-adiabatic transitions for $\lambda=4$ and
$\eta=50$ as a function of $\delta$. 
\label{numerical2}}
\end{center}
\end{figure}
\begin{figure}[htb]
\begin{center}
\includegraphics[width=0.8\textwidth]{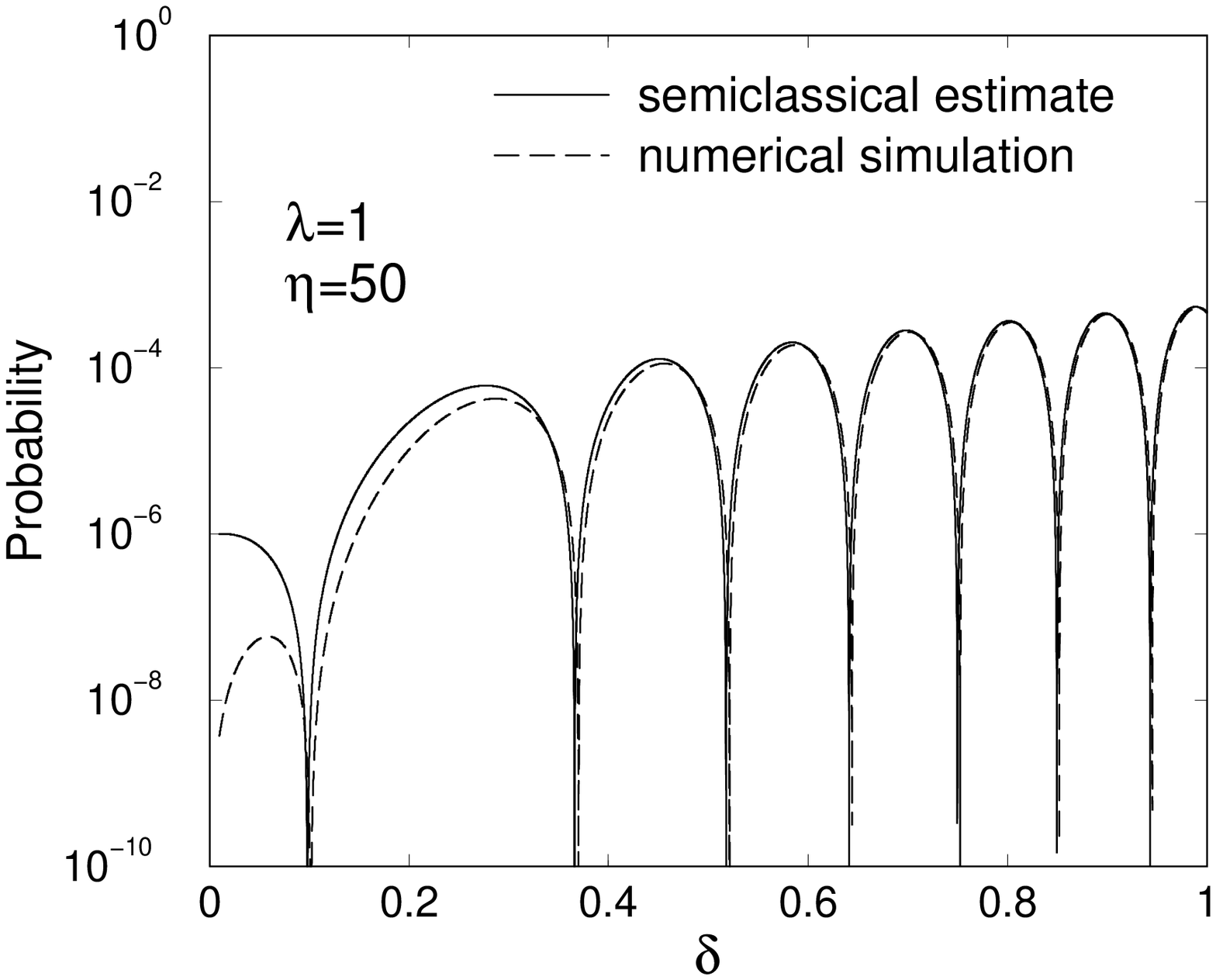}
\caption{The probability for non-adiabatic transitions for $\lambda=1$ and
$\eta=50$ as a function of $\delta$. 
\label{numerical3}}
\end{center}
\end{figure}
In Figures~\ref{numerical1}, \ref{numerical2}, and \ref{numerical3}, we 
compare our semiclassical estimate (\ref{main}) with results from 
numerical simulations of the exact quantum mechanical time evolution, 
following Ref.~\cite{Schliemann01a}. Both results are in  
excellent agreement and differ only at very small $\delta$, i.e.
for weak pulses. Of course, the non-adiabatic transition probability 
vanishes in this limit, whereas the semiclassical approximation breaks 
down (see Section 3.3). This regime is beyond the exponential accuracy
of Dykhne's formula. The integrability of our model allows us to make 
precise statements about the form and magnitude of the corrections to 
Dykhne's formula (see Section 4). For example, in the limit 
$\lambda \gg 1$ and $\delta \lambda \ll 1$ we have from the 
expansion (\ref{expansion2}) that
$P_< \sim (2\pi\delta\lambda)^2 \; e^{-4\pi\lambda}$, 
while in the same limit the result of Dykhne's formula (\ref{main}) 
gives only the exponential factor $e^{-4\pi\lambda}$ without 
information about the prefactor.
This explains a trend among Figures~\ref{numerical1}, 
\ref{numerical2}, and \ref{numerical3}, namely the increasing 
range, in terms of $\delta$, of validity of Dykhne's formula with 
increasing $\lambda$.  
The value for the adiabaticity parameter $\lambda=2$, represented in 
Figure~\ref{numerical1}, corresponds to a ramp time
$\tau=4\hbar/U_{H}$, which was identified
in Ref.~\cite{Schliemann01a} as a practical lower bound to ensure
sufficient adiabatic behavior in a gate operation between two
quantum-dot spin-qubits. It is interesting that Dykhne's formula 
remains accurate for smaller values of $\lambda$, in particular 
$\lambda=1$ as seen in Figure~\ref{numerical3}.  The reason is that 
the results (\ref{main}) and (\ref{ex})
have an incidental factor of $2\pi$ in the exponent, giving in 
practical terms the requirement for exponential suppression
$2\pi\lambda \gg 1$.

The expressions (\ref{main2}) and (\ref{main}), along with 
Figures \ref{numerical1}, \ref{numerical2}, \ref{numerical3} 
comprise our main results. For 
the remainder of this section we will address the justification
and limitations of these results. 

\subsection{Origin of Dykhne's Formula}

\begin{figure}[htb]
\begin{center}
\includegraphics[width=0.8\textwidth]{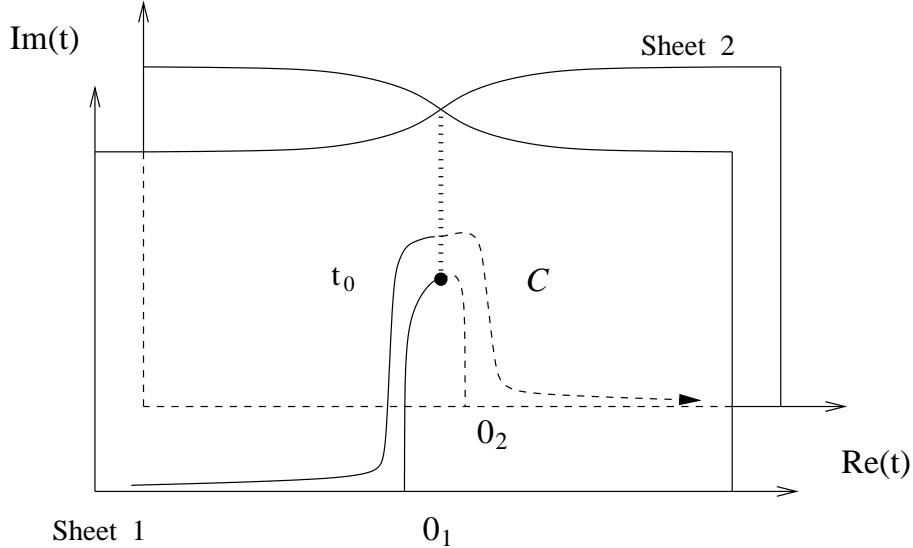}
\caption{An example of a Riemann surface with two 
sheets, a branch point at $t=t_0$, and a contour
$\mathcal{C}$ corresponding to a transition.}
\label{sheet}
\end{center}
\end{figure}

Dykhne derived a concise expression for non-adiabatic
transitions from a local analysis of the Schr\"{o}dinger equation
in the vicinity of the turning point \cite{Dykhne62}.
Dykhne's formula can be viewed as a semiclassical approximation, 
and an elegant interpretation and proof was given by Hwang and 
Pechukas~\cite{Hwang77} (see also \cite{Wilczek89}.) 
We will briefly discuss the key elements and scope of 
the proof.  Their method was to study
the solution of the Schr\"{o}dinger equation in the 
complex plane of the independent variable, time. 
According to the adiabatic theorem, the
projection of the solution onto any eigenstate
other than the initial eigenstate approaches zero 
in the adiabatic limit. One might suppose 
that weak statement is all the adiabatic theorem can 
tell us about transitions; however, it does
not exhaust its capacities. 
The reason lies in the following:
a basis of eigenstates $\Lb\xi_{1,2}(t)\Rr$, when
extended into the complex time plane, is multi-valued.
In particular, as a basis state is analytically continued 
across a branch cut of the function $\vep=\vep(t)$, 
its long time asymptotics are discontinuously changed. 
In accord with our above two-level problem, we uniquely
specify the basis by its asymptotics 
\ba
\Lb \xi_{1,2}(t) \Rr \ra \Lb S_{1,2} \Rr \qquad 
\textrm{as} \qquad t \ra \pm \infty .
\ea
The multi-valued nature is not manifest on the real time 
axis, because owing to the nondegeneracy of the 
spectrum $\pm \vep(t)$, the branch points are nonreal. 
We can choose a single-valued basis 
$\Lb\tilde{\xi}_{1,2}(t) \Rr$, which makes reference to 
$\Lb\xi_{1,2}(t)\Rr$ but has fixed asymptotics, 
by defining rules for continuing the basis states
across branch cuts.  Equivalently, this new basis is 
said to be defined over a Riemann surface 
with two sheets (copies of the complex time plane), 
one corresponding to each of the two branches of the 
function $\left[ \vep(t)^2 \right]^{1/2}$.  Crossing a 
branch cut means passing to the other sheet of 
the Riemann surface.  We assign the following 
relations among the eigenstates
\ba
\bar{lll}
\Lb \tilde{\xi}_{1,2}(t) \Rr = \Lb \xi_{1,2}(t) \Rr 
& \;\; t \in \textrm{sheet 1} & \textrm{and} \\
\Lb \tilde{\xi}_{1,2}(t) \Rr =
e^{\;\mi\; \alpha_{1,2}} \Lb \xi_{2,1}(t) \Rr  
& \;\; t \in \textrm{sheet 2} & \ear
\label{basis}
\ea
where $\alpha_{1,2}$ are phase conventions that are
chosen to maintain continuity of the basis 
$\Lb \tilde{\xi}_{1,2}\Rr$ across the branch cut (no 
physical quantity will depend on $\alpha_{1,2}$.)
Given $\Lb \psi(-\infty) \Rr = \Lb \tilde{\xi}_{1}(-\infty) \Rr$, 
the conclusion of the adiabatic theorem may be restated 
on a Riemann surface as 
\ba
\Lb \Ll \tilde{\xi}_{1}(t) | \psi(t) \Rr \Rb &\ra& 1 \qquad 
\qquad  \forall \;t \qquad \mathrm{as} \qquad  \tau \ra \infty ,
\label{Riemann}
\ea
where $\tau$ is the characteristic time scale for variation
of $\mathcal{H}_{\mathrm{eff}}(t)$. The only exception to
(\ref{Riemann}) is for times within $\tau \mathcal{O}(\lambda^{-2/3})$
of a turning point, for there the semiclassical criterion
(\ref{semiclass}) is invalid.
As remarked above, the zeroth order approximation 
(\ref{zeroth}) of the solution as $\tau \ra \infty$ is the 
dynamical phase. The zeroth order approximation may
be extended into the complex plane by evaluating the
dynamical phase on a contour $\mathcal{C}$. 
Continuing with the above example, a state that is purely
singly occupied at $t=-\infty$ is for complex time given by
\ba
\Lb \psi(t) \Rr \approx 
e^{-\mi \int_{\mathcal{C}} \mathrm{d}z \; (-\vep(z))} \; 
\Lb \tilde{\xi}_1(t) \Rr ,
\ea
where $\mathcal{C}$ is a contour from $z=-\infty$ to $z=t$.
The amplitude of transition is readily obtained as the projection
of the solution, as $t \ra \infty$ on the second
Riemann sheet (see Figure \ref{sheet}), onto the 
doubly occupied state $\Lb S_2 \Rr$, i.e.
\ba
\Ll S_2 | \psi(\infty) \Rr 
= \Ll \tilde{\xi}_{1}(\infty) | \psi(\infty) \Rr
\approx e^{\;\mi \alpha_1}\;
e^{-\mi \int_{\mathcal{C}} \mathrm{d}z \; (-\vep(z))} ,
\ea
where the contour $\mathcal{C}$ crosses the branch cut
emanating from the branch point
that is closest to the real axis. 
Dykhne's formula is simply the square 
modulus of this amplitude.

In the adiabatic regime, in contrast to the perturbative
regime, the leading contribution to transitions comes
from the zeroth order term of perturbation theory 
instead of the first order term.
By retaining only the zeroth order term, it appears
that we have neglected completely the coupling $t_H$ 
between states.  
However, the coupling enters implicitly in the multivalued 
function $\left[ \vep(t)^2 \right]^{1/2}$ and influences
the location of the turning points --- the complex roots 
of $\vep(t)$.  Transition amplitudes are 
obtained by carefully considering the different branches of 
this function. In the next section, we consider the validity
of keeping only the zeroth order term.

\subsection{Validity and Accuracy of Dykhne's Formula}

The theory of semiclassical approximations,
especially WKB analysis, provides a foundation from which to 
evaluate the validity of Dykhne's formula. The calculation of 
non-adiabatic transitions is closely related to the 
semiclassical approximation \cite{Hwang77}, because the 
semiclassical limit $\hbar \ra 0$ can be mathematically 
equivalent to the adiabatic limit $\tau \ra \infty$.
An essential element of the proof by Hwang and Pechukas is
the existence of a complex time contour that \mbox{}
1) connects the two sheets of the Riemann surface and \mbox{}
2) on which the zeroth order approximation of adiabatic 
perturbation theory is the correct leading behavior of the 
solution in the adiabatic limit.
These are sufficient conditions for Dykhne's formula to 
give the correct asymptotic form of the transition 
probability in the adiabatic limit $\lambda \ra \infty$.
Having established the existence of such a contour,
one can calculate a more precise value for the prefactor of 
Dykhne's formula by applying time dependent perturbation 
theory along the contour. We expect Dykhne's formula to 
breakdown when the contour ceases to exist. 
At the limit of its range of validity, the higher order terms 
become comparable to the zeroth order term. 
Introducing the unitary transformation $U$ that diagonalizes the
Hamiltonian, i.e. $U^{\dag} H U = \vep \sigma_3$, we can write the 
Schr\"{o}dinger equation in the basis of instantaneous eigenstates
\ba
\mi\,\hbar \, \frac{\mathrm{d}}{\mathrm{d}t} \; \Lb\xi(t) \Rr = 
\LB \vep(t) \;\sigma_3\; + \hbar \hat{a}(t) \RB  \Lb \xi(t) \Rr
\ea
with the off-diagonal perturbation  
$\hat{a}(t) = - U^{\dag} \;\mi \, \de_t \;U$.
A dominancy balance among the terms gives the condition for
the accuracy of the zeroth order approximation 
\ba
\frac{\Lb \vep(t) \Rb}{\hbar} \gg \Lb \hat{a}(t) \Rb
\ea
or in scaled time $x=t/\tau$,
\ba
\frac{\Lb \vep(x) \Rb \tau}{\hbar} \gg \Lb \hat{a}(x) \Rb . \label{balance}
\ea
For our model of the dynamics, $\lambda \propto \tau$ is 
the largest scale and $\Lb \hat{a} \Rb \sim 1$. 
The condition (\ref{balance}) must be maintained at all 
points on the contour.  Applying (\ref{balance}) on the
real axis, where $\Lb \vep(x) \Rb \sim U_H$, gives 
the adiabaticity condition $\lambda \gg 1$. 
Additionally, in order to connect two Riemann sheets,  
the contour must pass between two turning points 
(see Figure \ref{contour}), where 
$\Lb \vep(x) \Rb \sim \delta U_H$, giving the condition 
$\delta \lambda \gg 1$. 

Beginning instead from an intuitive approach, we can evaluate
the adiabaticity of the dynamics along a given contour.
To test whether a given contour is adequate, we can exploit the
analogy between quasi-adiabatic dynamics and semiclassical 
scattering. Recall the semiclassical criterion
\ba
\frac{\Delta (\Lambda)}{\Lambda} 
\sim \frac{\mathrm{d}\Lambda}{\mathrm{d}t} \ll 1 . \label{semiclass}
\ea
The analog of the de~Broglie wavelength $\lambda(x) = 2\pi\hbar/p(x)$ in 
scattering problems is the period $\Lambda(t) \equiv 2\pi\hbar/\vep(t)$. 
In other words, the condition (\ref{semiclass}) says that the change 
of period over the course of one period is small. 
We now require that the semiclassical criterion be obeyed 
{\em everywhere} along an admissible contour, i.e. one that
connects the two Riemann sheets.
To find an admissible contour, we must
appeal to the analytic structure of the eigenenergy $\vep=\vep(t)$,
see Figure \ref{contour}.
For clarity we will focus on the time interval $t<0$ and 
operate under the approximation (\ref{time}). The singularities of
$\vep(t)$ are branch points at 
$t= -\frac{T}{2} - \tau \;\ln(-1\pm\mi \delta)$
and poles at $t=-\frac{T}{2}- \tau \;\ln(-1)$. If we agree to define
a branch cut connecting the nearest and next nearest branch points
to the real time axis 
\ba
t_1&=&-\frac{T}{2}-\tau \;\ln \sqrt{1+\delta^2} 
+ \mi \tau \;\LB \pi- \arctan(\delta) \RB \qquad \mathrm{and} \nonum\\ 
t_3&=&-\frac{T}{2}-\tau \;\ln \sqrt{1+\delta^2} 
+ \mi \tau \;\LB \pi+ \arctan(\delta) \RB , \nonum
\ea
respectively, 
then an admissible contour is one that crosses this branch cut exactly
once.
For the semiclassical criterion to be obeyed, the 
admissible contour cannot pass too close to a branch point.
In essence, if $\delta$ is too small, the contour is pinched 
between the branch points $t_1$ and $t_3$. 
Evaluating the maximum of $\mathrm{d}\Lambda/\mathrm{d}t$ 
on a contour $\mathcal{C}$ that crosses the branch cut between
$t=t_1$ and $t=t_3$ we arrive at the condition
$\delta \gg \lambda^{-1}$. 
Together with the adiabatic limit $\lambda \gg 1$, we have the
following conditions on the interdependence of the physical
parameters
\ba
\lambda \sim \frac{U_H \tau}{\hbar} &\gg& 1 \qquad \mathrm{and} 
\label{ineq1} \\
\delta \lambda \sim \frac{\delta U_H \tau}{\hbar} &\gg& 1 . \label{ineq2}
\ea
Each of these dimensionless quantities is a product of a 
characteristic energy and time scale. 
If these conditions are not satisfied, there does not exist
a contour on which the motion is adiabatic.  The integrability
(Section 4) of our model allows us to investigate the intermediate 
regime $\lambda \gg 1$ and $\delta \lambda \ll 1$, where
Dykhne's formula cannot be justified with the analysis
of Hwang and Pechukas.

\section{Identification with an Integrable Model}

The result obtained by Dykhne's formula in Section 3.1 is 
now shown to be equivalent to the exact result 
for an integrable model in the appropriate limit.

Under the approximations (\ref{time}) for the time intervals
$t<0$ and $t>0$, the Hamiltonian 
\ba
\mathcal{H}_{\mathrm{eff}} = - \frac{U_H}{2}
\left( \begin{array}{cc} 1 & t_H \\
    t_H & -1 \end{array}\right) 
\ea
is approximated by
\ba
\mathcal{H}_{\mathrm{eff}} \approx
\left\{ 
\bar{lc}
\mathcal{H}_< & \qquad t<0\\[0.1cm]
\mathcal{H}_> & \qquad t>0
\ear \right. \label{timepatch}
\ea
with
\ba
\mathcal{H}_< &=&  - \frac{U_H}{2} \left( \begin{array}{cc} 
 1 & \frac{\delta}{1+ e^{-\frac{t}{\tau}-\frac{T}{2\tau}}} \\
 \frac{\delta}{1+ e^{-\frac{t}{\tau}-\frac{T}{2\tau}}}  & -1 
\end{array} \right) \\[0.5cm]
\mathcal{H}_> &=&  - \frac{U_H}{2}\left( \begin{array}{cc} 
 1 & \frac{\delta}{1+ e^{\frac{t}{\tau}-\frac{T}{2\tau}}} \\
 \frac{\delta}{1+ e^{\frac{t}{\tau}-\frac{T}{2\tau}}}  & -1 
\end{array} \right).
\ea
The Hamiltonians $\mathcal{H}_{<}$ and $\mathcal{H}_{>}$ can
be obtained as a special case of
\ba
\mathcal{H}_{\mathrm{exact}}= 
\left( \begin{array}{cc} b & a+c \;\tanh \frac{x}{2} \\
    a+c \; \tanh \frac{x}{2} & -b\end{array}\right) 
\ea
by identifying $\pm c = a = -\delta\lambda/2$, $b=-\lambda$ and 
rescaling time 
$x=t/\tau \pm T/2\tau$, respectively.
The Schr\"{o}dinger equation
\ba
\mi \de_x \Lb \psi(x) \Rr = \mathcal{H}_{\mathrm{exact}} \Lb \psi(x) \Rr
\ea
is exactly solvable \cite{Narozhny70} (see Appendix.) 

\begin{figure}[htb]
\begin{center}
\includegraphics[width=0.8\textwidth]{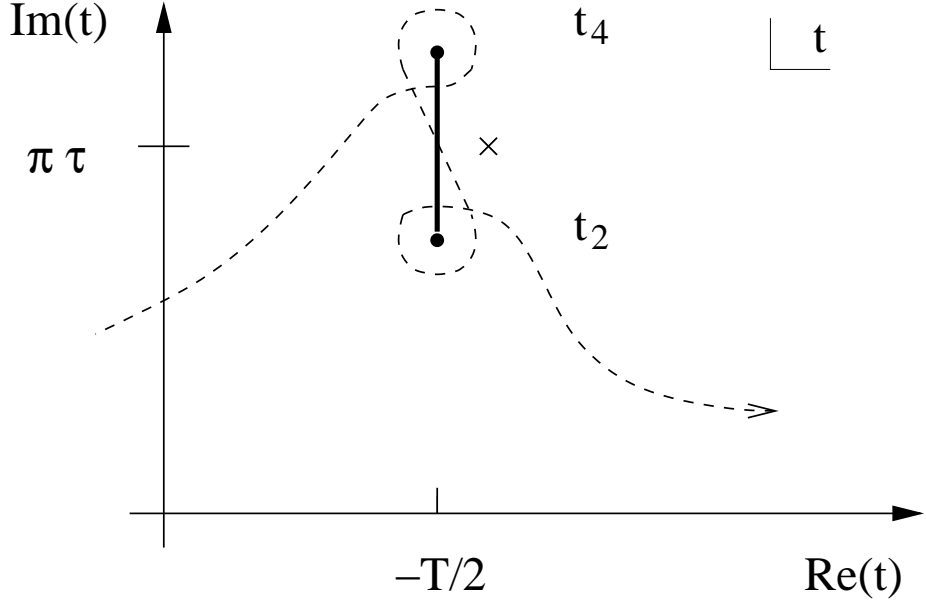}
\caption{The contour $\mathcal{C}$ gives a subdominant
correction to Dykhne's formula.}
\label{bubble}
\end{center}
\end{figure}

In analogy with one-dimensional scattering, the transition 
amplitude from a singly occupied state $\Lb S_1 \Rr$ to a 
doubly occupied state $\Lb S_2 \Rr$ may be viewed as an 
off-diagonal element of the scattering matrix $\mathcal{S}$ that 
connects the coefficients of the asymptotic final states 
to the asymptotic initial states. The asymptotic states
are the limit as $t\ra\pm\infty$ of the instantaneous 
eigenstates $\Lb \xi_{1,2}(t) \Rr$ of $\mathcal{H}_{\mathrm{eff}}$
corresponding to eigenvalues $\mp \vep(t)$, respectively,
\ba
\Lb \xi_1(t)\Rr &=& 
\frac{1}{\sqrt{2\vep}} \LB \bar{r} -\sqrt{\vep-\lambda} \\ 
\sqrt{\vep+\lambda} \ear \RB
\qquad \qquad \mathrm{and} \nonum\\[0.7cm]
\Lb \xi_2(t)\Rr &=& 
\frac{1}{\sqrt{2\vep}} \LB \bar{r} \sqrt{\vep+\lambda} \\ 
\sqrt{\vep-\lambda} \ear \RB . \label{estate}
\ea
The leading behavior of the long time asymptotics of a general 
solution has the form
\ba
\Lb \psi(t) \Rr = 
\left\{ \bar{ll} 
a_1 \; e^{\;-i \int^t \;\mathrm{d}t\pr \; \LB -\vep(t\pr)\RB}\; 
\Lb \xi_1(t) \Rr 
+ a_2 \; e^{\;-i \int^t \;\mathrm{d}t\pr \;  \vep(t\pr)}\; 
\Lb \xi_2(t) \Rr & \;\textrm{as}\;\; t \ra -\infty \\[0.5cm]
 b_1 \; e^{\;-i \int^t \;\mathrm{d}t\pr \; \LB -\vep(t\pr)\RB}\; 
\Lb \xi_1(t) \Rr 
+ b_2 \; e^{\;-i \int^t \;\mathrm{d}t\pr \; \vep(t\pr)}\; 
\Lb \xi_2(t) \Rr & \;\textrm{as}\;\; t \ra \infty \ear \right.
\ea
and the scattering matrix relates the coefficients 
\ba
\LB \bar{c} b_1  \\  b_2 \ear \RB = \mathcal{S} \; 
\LB \bar{c} a_1  \\  a_2 \ear \RB . \label{scatt}
\ea
Referring to the statement of our problem in Section 3, 
the amplitude of final double occupancy is the element $\mathcal{S}_{21}$
of the scattering matrix, which we parametrize as 
\ba
\mathcal{S} = \LB \bar{cc} \mathcal{S}_{11} & \mathcal{S}_{12} \\ 
-\mathcal{S}_{12}^* & \mathcal{S}_{11}^* \ear \RB .
\ea
The scattering matrices $\mathcal{S}_<$ and $\mathcal{S}_>$ associated with the 
Hamiltonians $\mathcal{H}_{<}$ and $\mathcal{H}_{>}$ may be
obtained by substitution from the exact scattering matrix 
$\mathcal{W}$ (derived in the Appendix) 
associated with the Hamiltonian $\mathcal{H}_{\mathrm{exact}}$.  
By the symmetry of the pulse, we have $\mathcal{S}_> = \mathcal{S}_<^{\dag}$. 
Patching together the two domains of time evolution (\ref{timepatch}),
we find the scattering matrix 
\ba
\mathcal{S} = \mathcal{S}_<^{\dag} \; \exp \LB \; \frac{\mi}{\hbar} \;\sigma_3 \;\Re \int_{t_1}^{t_2} \; \mathrm{d}t \;\vep(t) \RB  \; \mathcal{S}_< ,
\ea
where the integral of the exponent has been estimated in (\ref{argument}),
and the elements of $\mathcal{S}_<$ are obtained from $\mathcal{W}$
\ba
\LB \mathcal{S}_< \RB_{11} &=& \sqrt{\frac{2\mu}{\mu+\lambda}} \;
\frac{\Gamma(\mi 2\lambda) \;\;\Gamma(\mi 2\mu)}{\Gamma(\mi \mu +\mi \lambda +\mi \delta\lambda) \; \Gamma(\mi \mu +\mi \lambda -\mi \delta\lambda)} , \\[0.5cm]
\LB \mathcal{S}_< \RB_{12} &=& \sqrt{\frac{2\mu}{\mu-\lambda}} \;
\frac{\Gamma(\mi 2\lambda) \;\;\Gamma(-\mi 2\mu)}{\Gamma(-\mi \mu + \mi \lambda  +\mi \delta\lambda) \; \Gamma(-\mi \mu + \mi \lambda  -\mi \delta\lambda)} 
\ea
where $\mu = \lambda \sqrt{1+\delta^2}$.
Dykhne's formula (\ref{main2}) for $P_<$ is recovered  
as exactly the leading term of $\Lb \LB \mathcal{S}_< \RB_{21} \Rb^2$ in 
the limit $\lambda, \delta \lambda \gg 1$,
\ba
P_< &=& \Lb \LB \mathcal{S}_{<}\RB_{21} \Rb^2 \nonum \\[0.5cm]
  &=& \frac{\sinh \left[\pi \lambda \LB\sqrt{1+\delta^2}-1+\delta\RB\right] \;\;\sinh \left[ -\pi \lambda \LB\sqrt{1+\delta^2}-1-\delta\RB\right]}{\sinh \LB 2\pi \lambda \RB \;\;\sinh \LB 2 \pi \lambda \sqrt{1+\delta^2}\RB} \label{ex} \\[0.5cm]
&\sim &  e^{-2\pi \lambda \LB 1+\sqrt{1+\delta^2}-\delta\RB} 
\LB 1 - e^{-2\pi \lambda \LB \delta + 1 - \sqrt{1+\delta^2} \RB } + ... \RB .
\label{expansion}
\ea
The nonperturbative corrections are typically very small.  
For our problem $\lambda = 2$, $\delta = 1/2$, and the 
relative contribution of the second term of (\ref{expansion}) 
is less than one percent. 
This accounts for the excellent agreement in Figures 
\ref{numerical1}, \ref{numerical2}, and 
\ref{numerical3}, between the probability of double 
occupancy as given
by the semiclassical result (\ref{main}) based on Dykhne's
formula and the result of a numerical integration of the 
Schr\"{o}dinger equation.
We can interpret the subleading term of (\ref{expansion}) as the
contribution from the contour $\mathcal{C}$ of Figure \ref{bubble},
which crosses the branch cut three times.
The sign of the correction is negative and arises from 
the factor $e^{\mi \alpha_1}\; e^{\mi \alpha_2}$ associated
with matching the basis (\ref{basis}) across the branch cut. 
For our problem $\alpha_1 =0$ and $\alpha_2=\pi$.
Similarly, it may be possible to extend Dykhne's formula and
reproduce the exact result (\ref{ex}) by summing over
all homology classes of the Riemann surface and within 
each homology class over all paths that
give distinct nonzero values for 
$\Im \int_{\mathcal{C}} \mathrm{d}t \; \vep(t)$ \cite{Lam96}.
For many physical problems this type of nonperturbative
correction is dominated by perturbative corrections 
along the contour --- those mentioned in Section 3.3.  The 
striking absence of perturbative corrections in the limit 
$\lambda, \delta\lambda \gg 1$ is a unique artifact of
the integrability of our model. 

With a knowledge of the exact result, we can also investigate
the intermediate regime $\lambda \gg 1$ and 
$\delta\lambda \ll 1$, where the analysis of Hwang and Pechukas
(Section 3.2) cannot be used to prove Dykhne's formula.
In this limit, the transition probability (\ref{ex}) becomes
\ba
P_< \sim \LB 2 \pi\delta\lambda \RB^2 \; e^{-4\pi\lambda} .
\label{expansion2}
\ea
Although Dykhne's formula does not apply in this limit, 
because $\delta^{-1}$ and not $\lambda$ is
the largest scale, it nevertheless gives the correct 
controlling factor $e^{-4\pi\lambda}$ of (\ref{expansion2}),
except for $\delta$ that are exponentially small with 
respect to $\lambda$.
This exponential factor is resilient and remains the
controlling factor for a range of parameters beyond
the naive expectation based on the arguments of 
Section 3.3.

\section{Conclusions}

The dynamics of two coupled quantum dot spin-qubits can be
mapped to an effective two-level system, where non-adiabatic
transitions correspond to double occupancy.
We have estimated the probability of final double occupancy
with Dykhne's formula. In the 
adiabatic regime, the pervasive feature of transitions 
is their exponential suppression by a dimensionless adiabaticity
parameter, $\lambda$.   
Our main result (\ref{main}) was expressed in terms of
the dimensionless quantities $\lambda$, $\delta$ and $\eta$. 
An integral constraint (\ref{constraint}) on the swapping 
operation gives one relation among the three dimensionless 
parameters. The problem is uniquely defined by specifying any
two, and in a solid state setting, conservative 
estimates are $\lambda \approx 2$ and $\delta\approx 1/2$.
The probability of double occupancy $P \approx 10^{-10}$ 
is sufficiently rare that the operation of a quantum gate 
will not be obstructed by this type of error. It is 
noteworthy that the probability of double occupancy 
(\ref{main}) has nodes for 
\ba
k \pi = \Re \int_{t_1}^{t_2} \; \mathrm{d}t \;\vep(t) \approx 
\sqrt{1+\delta^2} \lambda \eta \qquad \qquad k \in \mathbf{Z} .
\ea
However, this property is not immediately relevant to the 
suppression of transitions, because the oscillatory factor, 
$\sin^2 \LB \sqrt{1+\delta^2} \lambda \eta \RB$, of (\ref{main})
vanishes algebraically and for it to provide an improvement
upon the exponentially small factor from Dykhne's formula,
the argument would have to be tuned exponentially close
to $k \pi$. Thus, naively the errors associated with inaccuracies
in satisfying the integral swapping constraint (\ref{constraint})
will be much greater than double occupancy errors. Other 
important sources of error are dephasing and decoherence of the
qubit states. 

We have reviewed a physically motivated derivation of Dykhne's 
formula \cite{Hwang77}.
The theory of semiclassical approximations underlies Dykhne's 
formula and its validity is appropriately judged within that 
framework. The semiclassical estimates obtained from this approach
are in excellent agreement with numerical simulations of the full
quantum mechanical time evolution.
The corrections to Dykhne's formula are of two types:
perturbative and nonperturbative.  The former appear to vanish 
for integrable models, and we have interpreted a 
nonperturbative correction as the contribution of a contour 
in the complex time plane \cite{Lam96}.

\section{Acknowledgments}

AGA and RR wish to thank Harold Metcalf for helpful 
discussions.
JS and DL acknowledge financial support from the Swiss
NSF, the NCCR Nanoscience, EU RTN Spintronics, DARPA, ARO, and ONR.
AGA received support from the Theory Institute for Strongly
Correlated and Complex Systems at BNL
as well as from NSF grant DMR-0348358.

\appendix

\section{Integrable Dynamics}

The following Hamiltonian~\cite{Narozhny70}, which 
possesses three independent parameters 
\ba
\mathcal{H}_{\mathrm{exact}}(x)
= \left( \begin{array}{cc} b & a+c \;\tanh \frac{x}{2} \\
    a+c \; \tanh \frac{x}{2} & -b\end{array}\right)  
\ea   
has integrable dynamics in the sense that the Schr\"{o}dinger equation,
\ba
\LB \mi\;\de_x - \mathcal{H}_{\mathrm{exact}}(x) \RB \; \Lb \psi(x) \Rr  
= 0 \qquad \textrm{with} \qquad
\Lb \psi(x) \Rr
= \left( \begin{array}{c} c_1(x) \\ c_2(x) \end{array} \right)
\ea 
has a solution in terms of special functions. Our aim is to 
construct the scattering matrix $\mathcal{W}$ associated with 
the dynamical problem. For clarity of presentation, we will consider 
the Hamiltonian 
\ba
\mathcal{H}\pr(x) = 
\left( \begin{array}{cc} a+c \;\tanh \frac{x}{2}  & b  \\
   b  & - \LB a+c \; \tanh \frac{x}{2} \RB \end{array}\right) ,
\ea
which differs from $\mathcal{H}_{\mathrm{exact}}$ by 
a constant unitary transformation $V$
\ba
V^{\dag} \; \mathcal{H}_{\mathrm{exact}} \;V 
= \mathcal{H}\pr \qquad \textrm{with} \qquad
V = \frac{1}{\sqrt{2}}\; \left( \begin{array}{cc} 1 & 1 \\
     1  & -1 \end{array}\right) .
\ea
Introducing the mixing angle $\phi$ defined as
\ba
\tan \phi = \frac{b}{a + c \;\tanh \frac{x}{2}}
\ea
the instantaneous eigenstates $\Lb\chi_{1,2}(x)\Rr$ corresponding to 
instantaneous eigenvalues
\ba
\pm \vep(x) = \pm \sqrt{b^2 + \LB a+c \tanh \frac{x}{2} \RB^2}
\ea
respectively, are parametrized as
\ba
\Lb \chi_1(t)\Rr &=& \frac{1}{\sqrt{2\vep}}
\LB \bar{c} \sqrt{\vep + (a+c \tanh \frac{x}{2})} \\ 
\sqrt{\vep - (a+c \tanh \frac{x}{2})} \ear \RB
= \LB \bar{c} \cos \frac{\phi}{2} \\ \sin \frac{\phi}{2} 
\ear \RB 
\qquad  \mathrm{and} \\[0.7cm]
\Lb \chi_2(t)\Rr &=& \frac{1}{\sqrt{2\vep}}
 \LB \bar{c} -\sqrt{\vep - (a+c \tanh \frac{x}{2})}  \\ 
 \sqrt{\vep + (a+c \tanh \frac{x}{2})} \ear \RB
= \LB \bar{c} -\sin \frac{\phi}{2} \\ \cos \frac{\phi}{2} 
\ear \RB . 
\ea

The two-state Schr\"{o}dinger equation may be converted 
to a second order differential equation for the components 
$c_{1,2}=c_{1,2}(x)$,
\ba
\de_x^2 c_{1,2} + 
Q_{1,2}(x) \;c_{1,2} = 0 
\label{differential}
\ea
where
\ba
Q_{1,2}(x) &=&
b^2 + \LB a + c \tanh \frac{x}{2}\RB^2 \pm 
\mi\;\frac{1}{2} \;c \;\; \mathrm{sech}^2 \frac{x}{2} .
\ea
Changing the dependent and independent variables as 
$c_{1,2}(x) = z^{\pm \mi \nu} \; \LB z-1 \RB^{\mi \mu} w_{1,2}(z)$ and
$z = \frac{1}{2} \LB 1+ \tanh \frac{x}{2} \RB$, transforms 
(\ref{differential}) into the standard form of the Gauss 
hypergeometric equation, for example, for $w_1(x)$
\ba
z (1-z) \;\de_z^2 w_1 + \left[k-z \LB i+j-1 \RB \right] \;\de_z w_1
-ij\; w_1 = 0 ,
\ea
where the arguments and exponents are 
defined as follows
\ba
i &=& \mi \mu + \mi \nu - \mi 2 c  \nonum\\
j &=& \mi \mu + \mi \nu + \mi 2 c +1 \nonum\\
k &=& \mi 2 \nu +1 \nonum
\ea
\ba
\nu &=&  \vep(-\infty) =  \sqrt{b^2+\LB a-c \RB^2} \nonum\\
\mu &=&  \vep(\infty)  =  \sqrt{b^2+\LB a+c \RB^2}. \nonum 
\ea
The two linearly independent solutions $u_1(z)$, $v_1(z)$ of the 
differential equation for $w_1(z)$ are
\ba
u_1(i,j,k;z) &=& z^{\mi \nu} \; \LB z-1 \RB^{\mi \mu}\;
  _2\mathrm{F}_1 \LB i,j,k;z \RB \qquad \textrm{and} \\[0.5cm]
v_1(l,m,n;z) &=& z^{1-k}\; u_1(i-k+1,j-k+1,2-k;z) \nonum\\[0.3cm]
&=& z^{-\mi\nu} \; \LB z-1 \RB^{\mi \mu}\;
 _2\mathrm{F}_1 \LB i-k+1,j-k+1,2-k;z \RB,
\ea
where $_2\mathrm{F}_1 \LB i,j,k;z \RB$ is the Gauss
hypergeometric function \cite{Grad}. 

The amplitude of transition may be viewed as the off-diagonal element 
of the scattering matrix $\mathcal{W}$ that connects the 
asymptotic final and initial states for $t \ra \pm \infty$ 
(see (\ref{scatt}).) 
The scattering matrix is parametrized as
\ba
\mathcal{W} = \LB\bar{cc}  \mathcal{W}_{11} & 
\mathcal{W}_{12} \\ 
- \mathcal{W}^*_{12}  &  \mathcal{W}^*_{11}  
\ear \RB ,
\ea
because it has the properties 
$\mathcal{W}^{\dag}\; \mathcal{W} = 1$ 
and $\mathrm{det}(\mathcal{W})=1$.  
From the asymptotics of $u_1, u_2, v_1,\ \textrm{and}\ v_2$ in the limits 
$z\ra 0$ and $z \ra 1$, we find
\ba
\mathcal{W}_{11}
&=& \sqrt{\frac{b^2+\LB \mu - \LB a+c \RB \RB^2}{b^2+\LB \nu - \LB a-c \RB \RB^2}} \;
\frac{\Gamma(1-\mi 2\nu) \;\Gamma(-\mi 2\mu)}{\Gamma\LB 1-\mi \LB \mu+\nu-2 c\RB \RB \; \Gamma\LB -\mi \LB \mu+\nu+2 c\RB \RB} \\[0.5cm]
\mathcal{W}_{12}
&=&  \sqrt{\frac{b^2+\LB \mu + \LB a+c \RB \RB^2}{b^2+\LB \nu - \LB a-c \RB \RB^2}}\;
\frac{\Gamma(1-\mi 2\nu) \;\Gamma(\mi 2\mu)}{\Gamma\LB \mi \LB \mu-\nu-2 c\RB \RB \; \Gamma\LB 1+\mi \LB \mu-\nu+2 c\RB \RB} 
\ea
and
\ba
\Lb \mathcal{W}_{11} \Rb^2 &=& 
\frac{\sinh \pi \LB 2c +\mu +\nu \RB \;\; \sinh \pi \LB -2c +\mu +\nu \RB}{\sinh 2\pi \mu \;\; \sinh 2 \pi \nu} \\[0.5cm]
\Lb \mathcal{W}_{12} \Rb^2 &=& 
\frac{\sinh \pi \LB 2c -\mu +\nu \RB \;\; \sinh \pi \LB 2c +\mu -\nu \RB}{\sinh 2\pi \mu \;\; \sinh  2 \pi \nu } .
\ea

\end{document}